\documentclass[11pt,twoside]{article}

\usepackage{asp2004}
\usepackage{epsf}
\usepackage{psfig}
\usepackage{lscape}
\usepackage{graphicx}
\usepackage{amsmath}
\usepackage{amssymb}

\newcommand{\ve}[1]{{\rm{\bf {#1}}}}

\newcommand{\obs}{_{\rm obs}}
\newcommand{\sol}{_{\rm solar}}

\begin{document}
\title{Magnetic field vector retrieval with HMI}   
\author{J.M.~Borrero, S.~Tomczyk, A.~Norton, T.~Darnell}   
\affil{High Altitude Observatory, National Corporation for
Atmospheric Research, 3450 Mitchell Lane 80301, Boulder 80301 Colorado USA.}    
\author{J.~Schou, P.~Scherrer, R.~Bush, Y.~Lui}
\affil{W.W.~Hansen Experimental Physics Lab, Stanford University, 445 Via Palou,
Stanford 94305 California USA}

\begin{abstract} 
The Helioseismic and Magnetic Imager (HMI), on board the Solar
Dynamics Observatory (SDO), will begin data acquisition in 2008.
It will provide the first full disk, high temporal cadence observations
of the full Stokes vector with a 0.5 arc sec pixel size. This will allow
for a continuous monitoring of the Solar magnetic field vector. 
HMI data will advance our understanding of the small
and large-scale magnetic field evolution, its relation to the solar and global
dynamic processes, coronal field extrapolations, flux emergence, magnetic helicity
and the nature of the polar magnetic fields. We summarize HMI's 
expected operation modes, focusing on the polarization cross-talk 
induced by the solar oscillations and how this affects the magnetic 
field vector determinations.
\end{abstract}

\section{HMI polarization modulation}%
The HMI camera will measure the full
Stokes vector at 5 wavelength positions
along the Fe I 6173 \AA~ line (Graham et al. 2002). 
The CCD has 4024$\times$4024 pixels (0.5''/pixel). 
Our goal is to record the full Stokes vector for the whole solar disk with a cadence of
80-120 seconds. At each wavelength position two 
consecutive linear combinations of the solar Stokes vector 
are measured with a time delay of 4 seconds:

\begin{equation}
\mathcal{\ve{O}}_{n \times 1} = \mathcal{M}_{n \times 4} \ve{I}\sol
\end{equation}

Three different modulation matrices are under consideration: Mod A, Mod B
and Mod C. The total time needed to scan the line in the four polarization
states across five wavelength positions is $20 n$ seconds
($n$ is the number of rows in the modulation matrix in Eq.~2-4) : 
80s for mod A-B and 120 s for C.

\begin{equation}
\mathcal{M}_A = \left( \begin{array}{rrrr}
1.000 & 0.810 & 0.000 &0.588 \\
1.000 & -0.810 & 0.000 &0.588 \\
1.000 & 0.000 & 0.810 &-0.588 \\
1.000 & 0.000 & -0.810 &-0.588 \end{array} \right)
\end{equation}

\clearpage
\begin{figure}[!ht]
\begin{center}
\includegraphics[width=12cm]{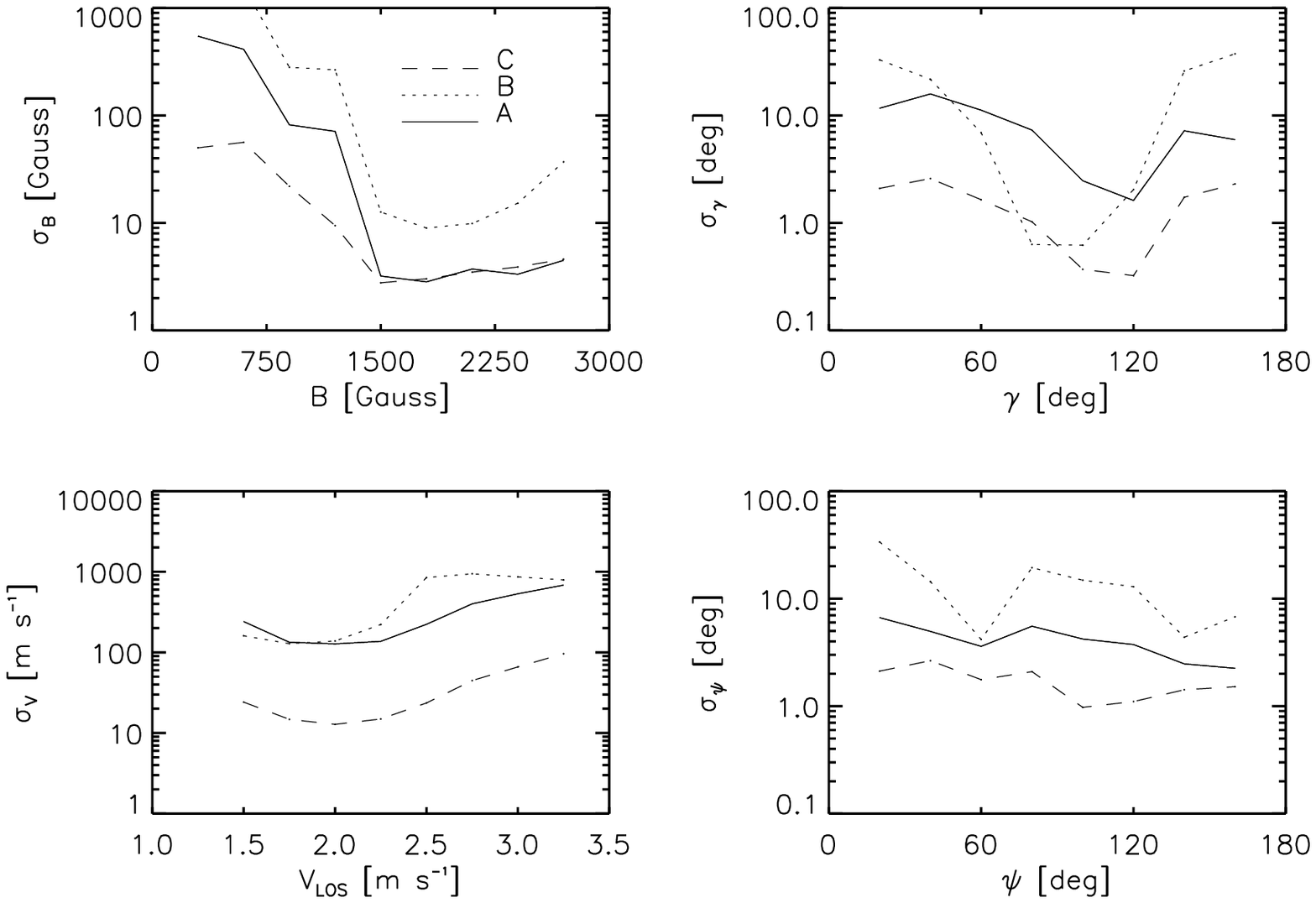} \\
\includegraphics[width=12cm]{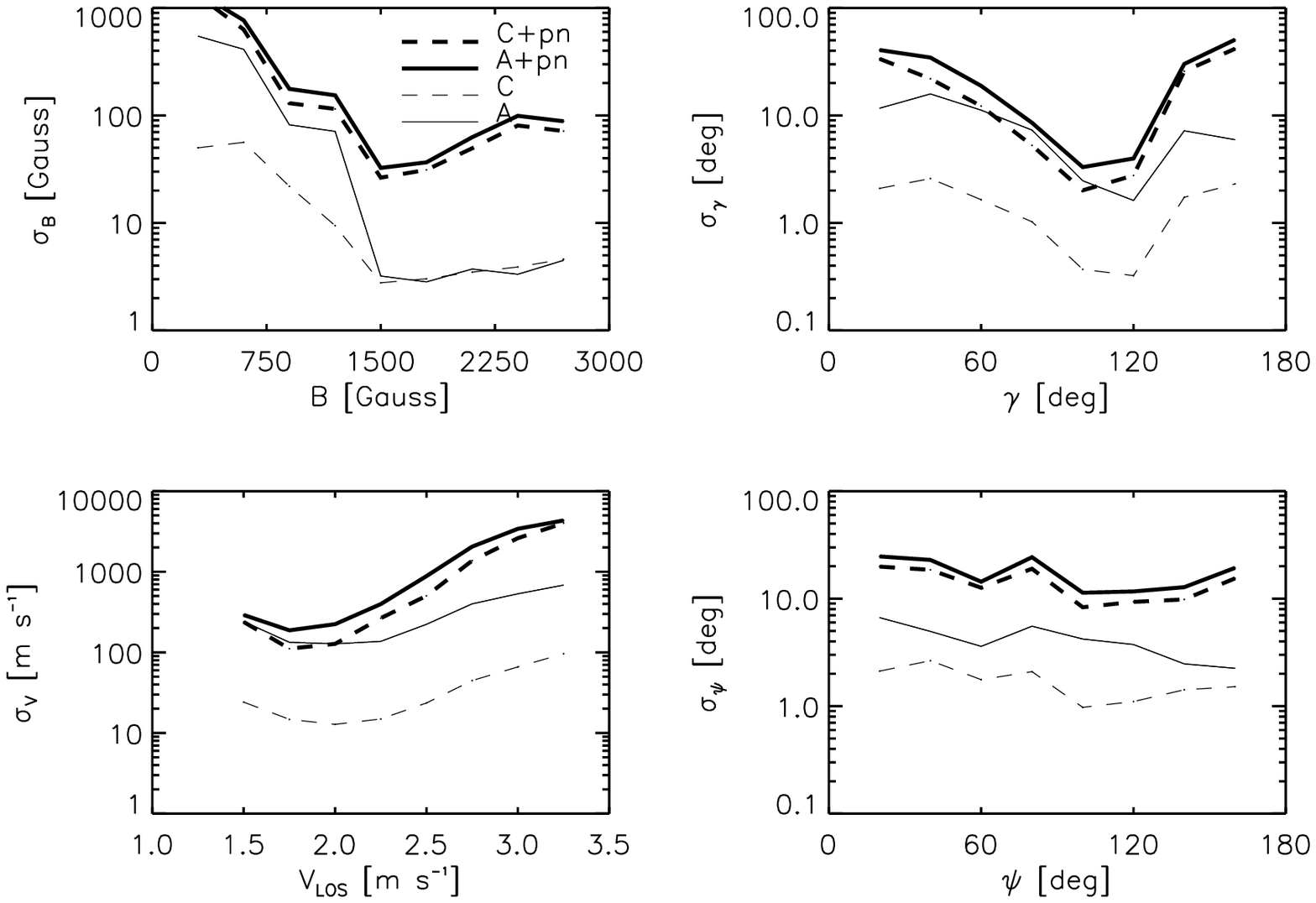}
\caption{{\it Top 4 panels}: errors in the determination of magnetic
field strength (upper left panel), field inclination
(upper right), field azimuth (bottom right) and
line of sight velocity (bottom left) as a consequence
of the solar p-modes using three different modulation
schemes: A (solid; Eq.~2), B (dotted; Eq.~3) and C (dashed;
Eq.~4). {\it Bottom 4 panels}: Same as above but using only mod A and mod C.
Thin lines correspond to results shown in upper four panel where only 
p-modes where considered. Thick lines
represent the errors with p-modes and photon noise.}
\end{center}
\end{figure}
\clearpage

\begin{equation}
\mathcal{M}_B = \left( \begin{array}{rrrr}
1.000 & 0.810 & 0.000 &0.588 \\
1.000 & 0.000 & 0.810 &-0.588 \\
1.000 & -0.810 & 0.000 &0.588 \\
1.000 & 0.000 & -0.810 &-0.588 \end{array} \right)
\end{equation}

\begin{equation}
\mathcal{M}_C = \left( \begin{array}{rrrr}
1 & 1 & 0 &0 \\
1 & -1 & 0 &0 \\
1 & 0 & 1 &0 \\
1 & 0 & -1 &0 \\
1 & 0 & 0 &1 \\
1 & 0 & 0 &-1 \end{array} \right)
\end{equation}

When the linear system is solved in
order to obtain the solar Stokes vector, observations
done at different times are mixed together. This introduces 
errors in the observed Stokes vector:

\begin{equation}
\ve{I}\obs = \mathcal{M}_{4 \times n}^{-1} \mathcal{\ve{O}}_{n \times 1} \ne 
\ve{I}\sol
\end{equation} 

For example, at $t=0$ modulation schemes measures
$I+0.81Q+0.59V$ and at $t=$4 seconds it measures:
$I-0.81Q+0.59V$. Therefore we can obtain stokes $Q$
by combining observations done 4 seconds apart.
Note that to build Stokes $V$ four different observations
(obtained at different times) are mixed.

\section{Effect of the Solar P-modes}%

To study the errors introduced in the determination of the 
solar magnetic field vector as a consequence of the solar p-modes,
we take them into account when computing $\ve{I}\obs$.
The different profiles (that enter into Eq.~5) are shifted in quantities 
that correspond to the velocity changes associated with the solar
p-modes. They are then inverted using
a Stokes inversion algorithm (Skumanich \& Lites 1987) that retrieves: 
the magnetic field vector, line of sight velocities and a filling factor 
(fractional area of the pixel covered by the magnetic atmosphere).
Results from the inversion are displayed in Fig.~1 (top panels).

\section{Photon noise}%

The noise level for a single observation using HMI
will be about $0.2 \times 10^{-3}$ in units of the continuum
intensity. This number considers a light level of the quiet Sun 
at disk center. For off-limb observations
and/or darker regions (sunspots) the noise level
will be different. A realistic simulation has been performed 
taking into account the solar p-modes effect and photon noise
simultaneously. In this case we restrict ourselves to
consider only modulation schemes A and C. Results are
presented in Fig.~1 (bottom panels).

\section{Temporal averages}%

Due to the oscillatory nature of the solar p-modes,
its effect on the observed profiles can be reduced
if we use time averaged profiles. By doing so we
can improve the accuracy in the determination
of the magnetic field vector. Note that time
averaging will also reduce the effect of the photon
noise.

We have carried out several simulations using different
averaging times. Fig.~2 displays the errors in
the determination of the magnetic field vector
as a function of the time used in the average. 
This was done both for Mod A and C. We
distinguish between different solar regions
by taking averages of pixels with different
filling factors.

\begin{figure}[!ht]
\begin{center}
\includegraphics[width=12.3cm]{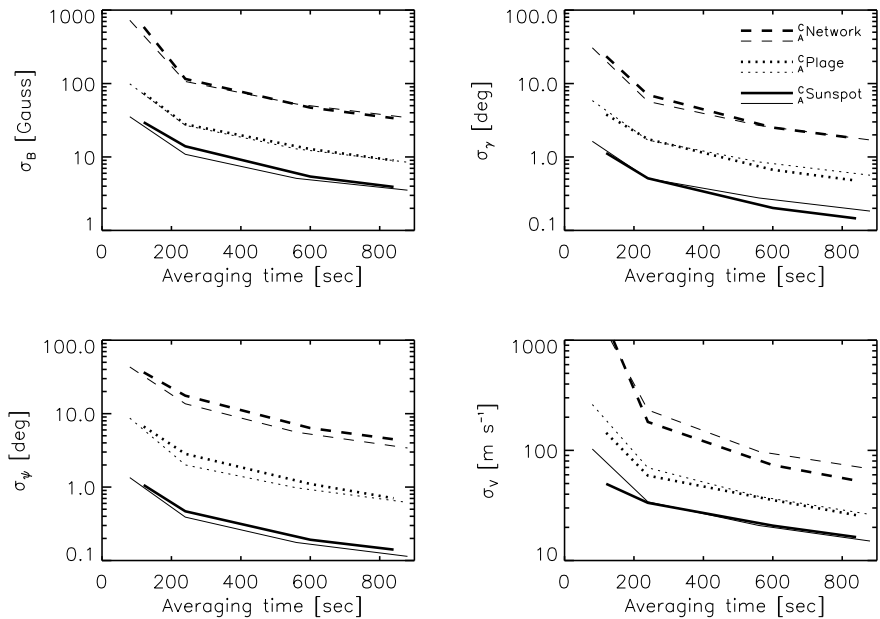}
\caption{Retrieved errors as a function
of the averaging time. Thin and thick lines correspond to 
Mod A and C respectively. Line styles indicate different 
solar regions: dashed (network), dotted (plages) and solid 
(sunspots).}
\end{center}
\end{figure}

\section{Conclusions}%

HMI will provide full disk observations of the magnetic field vector
with a cadence of 80-120 s and a spatial resolution of 0.5 arc sec.
We have considered the main sources of error in the 
data and estimate how this will affect the reliability of the inferred
vector. Accuracy can be improved by using time averaged observations. For a 
10 minutes average, the magnetic field in sunspots and plages can be known with 
a precision better than 10 Gauss and 1$^{\circ}$ in inclination and azimuth.


\end{document}